\documentclass[copyright,creativecommons]{eptcs}
\usepackage{breakurl}             

\usepackage{multicol}
\usepackage{mathptmx}
\usepackage{color}
\usepackage{amsmath}
\usepackage{amsthm}
\usepackage{amssymb}
\usepackage{stmaryrd}
\usepackage{drl-common/proof}
\usepackage{drl-common/typesit}
\usepackage{drl-common/typescommon}
\usepackage[square,numbers,sort]{natbib}
\usepackage{arydshln}
\usepackage{graphics}
\usepackage{natbib}

\usepackage{ucs}
\usepackage[utf8x]{inputenc}
\usepackage[T1]{fontenc}

\newcommand{\noagda}[0]{}

\usepackage{graphicx}
\newcommand\Shamrock{\includegraphics[width=0.6em]{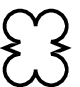}}

\DeclareUnicodeCharacter{8594}{$\shortrightarrow$}
\DeclareUnicodeCharacter{9001}{$\langle$}
\DeclareUnicodeCharacter{9002}{$\rangle$}
\DeclareUnicodeCharacter{12314}{$\llbracket$}
\DeclareUnicodeCharacter{12315}{$\rrbracket$}
\DeclareUnicodeCharacter{8872}{$\vDash$}
\DeclareUnicodeCharacter{9711}{\ensuremath{\bigcirc}}
\DeclareUnicodeCharacter{9675}{$\circ$}
\DeclareUnicodeCharacter{9671}{$\Diamond$}
\DeclareUnicodeCharacter{9657}{$\triangleright$}
\DeclareUnicodeCharacter{8640}{$\rightharpoonup$}
\DeclareUnicodeCharacter{8659}{$\Downarrow$}
\DeclareUnicodeCharacter{8984}{$\Shamrock$}

\usepackage{fancyvrb}

\newcommand{\ttt}[1]{\texttt{#1}}

\usepackage{drl-common/code}
\DefineVerbatimEnvironment{code}{Verbatim}{fontsize=\small,fontfamily=tt}

\newcommand{\ignore}[1]{}

\title{A Monadic Formalization of ML5}
\author{Daniel R. Licata\footnote{%
This research was sponsored in part by the National Science
Foundation under grant number CCF-0702381 and by the 
Pradeep Sindhu Computer Science Fellowship. The views and conclusions
contained in this document are those of the author and should not be
interpreted as representing the official policies, either expressed or
implied, of any sponsoring institution, the U.S. government or any other
entity.
}
\qquad
Robert Harper$^*$
\institute{Carnegie Mellon University}
\email{\{drl,rwh\}@cs.cmu.edu}}
\def\titlerunning{A Monadic Formalization of ML5}
\def\authorrunning{D.~Licata \& R.~Harper}

\begin{document}
\maketitle

\begin{abstract}
  ML5 is a programming language for spatially distributed computing,
  based on a Curry-Howard correspondence with the modal logic S5.
  However, the ML5 programming language differs from the logic in
  several ways.  In this paper, we give a semantic embedding of ML5 into
  the dependently typed programming language Agda, which both explains
  these discrepancies between ML5 and S5 and suggests some
  simplifications and generalizations of the language.  Our embedding
  translates ML5 into a slightly different logic: intuitionistic S5
  extended with a lax modality that encapsulates effectful computations
  in a monad.  Rather than formalizing lax S5 as a proof theory, we
  \emph{embed} it as a universe within the the dependently typed host
  language, with the universe elimination given by implementing the
  modal logic's Kripke semantics.
\end{abstract}

\section{Introduction}

One of the many benefits of formalizing programming languages and logics
is that the process of formalization, and the constraints of the
particular techniques used, can lead to new insights about the system
being studied.  This paper provides a worked example of this phenomenon,
investigating the ML5 programming language for spatially distributed
computing~\citep{murphy08thesis}.  ML5 has previously been
formalized~\citep{murphy08thesis} using syntactic methods in
Twelf~\citep{pfenningschurmann99twelf}.  However, we wished to give a
semantic interpretation of ML5 into a dependently typed programming
language, as a first step towards extending work on embeddings of
security typed-languages~\citep{ml10sectyp} to account for spatially
distributed access control, as in the PCML5 extension of
ML5~\citep{avijit10pcml5}.  Our semantic formalization of ML5 provides
insight into several discrepancies between ML5 and the logic upon which
it is based, and suggests some simplifications and generalizations of
the language, as we now describe.  

ML5 facilitates distributed programs that deal with \emph{located
  resources}, such as a database on a server, the browser display on a
client, or heap references on any particular site.  When a distributed
program running at one site attempts to access a resource located at
another site, the program must either communicate with the other site or
fail.  Because tacit communication makes it very difficult to reason
about the execution time of a program (e.g. every memory dereference
might involve a network communication), ML5 is based on the stance that
all communication should be explicit in the program.  However, rather
than letting accesses from the wrong site fail dynamically, ML5 employs
a type system based on a Curry-Howard correspondence with the modal
logic S5 to catch these errors statically.  ML5 is defined as an
intuitionistic modal logic in the style of
Simpson~\citep{simpson93thesis}, where hypotheses and conclusions are
considered relative to \emph{worlds}, which represent places on a
network. The ML5 typing judgement has the form $x_1 : A_1[w_1] , \ldots,
x_n : A_n[w_n] \vdash e : C[w]$, where $A_i$ and $C$ are modal types and
$w_i$ and $w$ are worlds.

Despite being designed by a correspondence with S5 modal logic, the ML5
programming language differs from S5 in several ways: First, ML5 ensures
that all communication is explicit in the program by providing only a
single communication primitive that, operationally, goes to world $w'$,
runs $e$, and brings the resulting value back to $w$:
\[
\infer{\Gamma \vdash \dsd{get}\ e : A [ w ]}
      {\Gamma \vdash e : A [ w' ] & A\ \dsd{mobile}}
\]
The condition $A\ \dsd{mobile}$ rules out
instances of \dsd{get} where $A$ is, for example, $\dsd{ref}\
\dsd{int}$---which would take a reference that should be used at $w'$
and turn it into a reference that should be used at $w$, violating the
intended typing guarantees.  

Second, in the standard presentation of S5, elimination rules for
positive connectives such as sums allow an arbitrary conclusion, which
is unconnected to the principal formula.  In ML5, this rule is
\emph{tethered}, in that the world in the conclusion must be equal to
the world in the premise:

\[
\infer[\text{untethered \dsd{case}}]{\Gamma \vdash C [ w' ]}
      {\begin{array}{l}
          \Gamma \vdash A \vee B [ w ] \\
          \Gamma , x:A[w] \vdash C [ w' ] \\
          \Gamma , x:B[w] \vdash C [ w' ]
        \end{array}}
\qquad \qquad
\infer[\text{tethered \dsd{case}}]{\Gamma \vdash C [ w ]}
      {\begin{array}{l}
          \Gamma \vdash A \vee B [ w ] \\
          \Gamma , x:A[w] \vdash C [ w ] \\
          \Gamma , x:B[w] \vdash C [ w ]
        \end{array}}
\]
ML5 makes the tethering restriction because the obvious operational
interpretation of the untethered rule requires communication (go to
\ttt{w} to run the principal formula).  Indeed, the untethered rule is
derivable using \ttt{get}.  However, this tethering is at odds with the
Kripke semantics of modal logic, where $A ∨ B [ w ]$ is interpreted as
$A [ w ]$ or $B [ w ]$---if the interpretation commutes with
disjunction, then a disjunction should be eliminable no matter the
conclusion of the sequent.

Third, ML5 includes two different $\Box$-like modalities with the same
introduction rule.  The first is written $⌘\ A$, while the second,
$\forall w.A~\dsd{at}~w$, is a composition of the connective $\forall$
(quantification over worlds) and the hybrid
logic~\citep{areces+01hybrid} \dsd{at} modality, which internalizes the
judgement $A [ w ]$ as a connective. (\emph{Hybrid
    logic} is between modal logic (truth is relativised to worlds)
and first-order logic (propositions may mention worlds)).
The two connectives are eliminated differently: ML5
distinguishes a syntactic category of values from ordinary expressions,
and $⌘~A$ can be eliminated to construct a value but $\forall
w.A~\dsd{at}~w$ cannot.

In addition to these discrepancies, there is some confusion over the
meaning the world in ML5 value judgements $v :: A [ w ]$ and expression
judgements $e : A [ w ]$.  An expression judgement means that $e$ is an
expression that must be evaluated at $w$, and produces a value $v :: A
[w ]$, but what does the world in the value judgement mean?  One cannot
think of values as just a subset of expressions, as the value rules for
certain connectives, such as \ttt{at} and ⌘, would violate the property
that all communication happens through \ttt{get}.  Additionally, 
in the dynamic semantics of the ML5 internal language given in Section
3.3 of \citet{murphy08thesis}, $\dsd{get}~e$ returns the entire value of
$e$ to the calling world, so the value judgement does not mean that the
value $v$ is physically located at $w$.

\medskip

In this paper, we propose a new logical foundation for ML5, which
explains the differences between ML5 and S5 and clarifies the role of
the world in a value judgement.  We translate ML5 into the
intuitionistic logic S5 extended with a lax modality, written $◯\ A$,
that encapsulates effectful computations in a
monad~\citep{moggi91monad,benton+98lax,curry52lax}.  This monadic
distinction between pure terms and effectful computations is already
tacit in ML5's distinction between values and expressions---and in
intermediate languages used in the ML5 compiler (e.g. the CPS language
in \citet{murphy08thesis}), which include elimination forms for
``values'' as ``values''.  Here, we draw out this distinction by
formalizing a monadic interpretation of ML5, and make some improvements
to the language based on this formulation.

In our interpretation, ML5 values of type $A[w]$ are interpreted as pure
terms of type $A^* \langle w \rangle$, where $A^*$ is a monadic
translation of $A$, and we write $B \langle w \rangle$ for a worlded
type in the lax modal logic.  On the other hand, potentially effectful
ML5 expressions are interpreted as inhabitants of type $(◯\ A^*) \langle
w \rangle$.  The role of the world in a value judgement, i.e. the role
of the world in $B \langle w \rangle$, is to describe where the
resources in subexpressions of the type $A$ may be used and where the
computations in the type $A$ must be run.  For example, the value
judgement $(\dsd{ref}~\dsd{int} ⊃ ◯ \dsd{unit}) \langle \dsd{client} \rangle$
describes a function that takes a reference that must be used at the
client and produces a computation that must be run at the client.



Our interpretation explains the three discrepancies between ML5 and S5
mentioned above: the \dsd{get} primitive is an extra operation on the
monad $◯$ that allows a computation that must be run at one world to be
run from another.  The tethered \ttt{case} rule is a derived rule: in
ML5, the scrutinee of the case is an effectful expression, so it is
necessary to sequence evaluating this expression with an actual case
analysis on the value produced---and it is the \emph{sequencing} that
requires tethering.  Indeed, we show that we can enrich ML5 with an
untethered case rule on \emph{values}, which would permit simpler code.
Finally, the $⌘$ connective can be eliminated in favor of $\forall$ and
\dsd{at}, given the standard pure elim rules for these types.

Rather than formalizing lax S5 as a proof theory, we \emph{embed} it
inside a dependently typed host language, Agda~\cite{norell07thesis}.
First, we define a lax logic for distributed programming, L5, which is
embedded in Agda using an indexed monad of computations at a place.
Next, we define a universe of hybrid modal types, HL5, and give them
meaning by interpretation into L5---i.e. we define a syntax of HL5
types, along with a function interpreting them as L5 types.  Finally, we
translate ML5 into HL5.  This technique saves us the work of defining a
proof theory for HL5, and additionally allows us to inherit the
equational theory of the meta-language, which can be exploited in
proving that the semantics validates the operational semantics of ML5.
While it is simple to embed type systems specified by standard
judgements of the form $x_1 : A_1 , \ldots, x_n : A_n \vdash e : C$ as a
universe, it requires a bit of thought to adapt these techniques to
languages with modal type systems, such as HL5.  In previous work on
programming with variable binding~\citep{lh09unibind}, we employed a
technique for embedding such modal type systems: intuitionistic modal
logics can be given a Kripke semantics in first-order intuitionistic
logic~\citep{simpson93thesis}, and we can formalize this semantics in a
dependently typed language host language.  However, the presentation of
this technique in that paper was somewhat obscured by the particular
example.  In this paper, we present this technique in a simpler setting,
and apply it to explain the proof theory of ML5.
The Agda code for this paper is available from
\verb|www.cs.cmu.edu/~drl|.  

\medskip
We briefly review Agda's syntax; see the Agda
Wiki\verb|(wiki.portal.chalmers.se/agda/)| for more introductory
materials.  Dependent function types are written with parentheses as
\ttt{(x : A) → B}. An implicit dependent function space is written as
\ttt{ \{x : A\} → B} or \ttt{∀ \{x\} → B} and arguments to implicit
functions are inferred.  Non-dependent functions are written as \ttt{A →
B}.  Functions are written as \ttt{λ x → e}.  Named functions are
defined by clausal pattern-matching definitions.  \ttt{Set} is the
classifier of classifiers in Agda, like the kind \ttt{type} in ML or
Haskell.

\ignore{
\begin{code}
open import lib.Prelude hiding ([_])

module agda5.Agda5 where

  postulate 
\end{code}}

\section{L5}

In this section, we define an interface for distributed programming in
Agda, based on lax logic~\citep{curry52lax,benton+98lax}.  We define a
type of worlds and a family of monads \ttt{IO w S} indexed by worlds,
which represent an effectful computation that runs at world \cd{w} and
produces a value of type \cd{S}.

\begin{code}
    World : Set 
    server : World
    client : World

    IO     : World → Set → Set
    return : ∀ {w S} → S → IO w S
    _>>=_    : ∀ {w S S'} → IO w S → (S → IO w S') → IO w S'
    get    : ∀ {w' w S} → IO w' S → IO w S

    Ref : World → Set → Set
    _:=_ : ∀ {w S} → Ref w S → S → IO w Unit
    !    : ∀ {w S} → Ref w S → IO w S
    new  : ∀ {w S} → IO w (Ref w S)
\end{code}
\ignore{
\begin{code}
    Run : ∀ {w S} → IO w S → S → Set
    Run/return  : ∀ {w S} {v : S} → Run{w} (return v) v
    Run/bind    : ∀ {w S S' e e'} {v : S} {v' : S'} → Run{w} e v → Run (e' v) v' → Run (e >>= e') v'
    Run/get     : ∀ {w w' S v} {e : IO w S} → Run e v → Run{w'} (get e) v
\end{code}}
The definitions are parametrized by a type of worlds, which we here
assume to contain client and server.  \cd{IO w S} is axiomatized as a
monad, with \cd{return} and bind (\cd{>>=}), along with an additional
operation \cd{get} that allows the computation to switch to a different
world.  The command \ttt{get} can be thought of as a remote procedure
call, which goes to \ttt{w'}, runs the given command, and then brings
the resulting value back to \ttt{w}.  Next, we define a type \ttt{Ref w
S} of heap references located at world \ttt{w}.  The operations for
setting (\ttt{:=}), getting (\verb|!|), and creating (\ttt{new})
references require the reference to be at the same world as the
computation.

Because these types are embedded in Agda, they may interact freely with
the host language types---e.g., a computation may return value of any
Agda \ttt{Set}, and we can use Agda Π and Σ types to quantify over
worlds.

In Agda, we use the \emph{postulate} keyword to assume an implementation
of this interface.  Under the hood, these operations can be implemented
using foreign-function calls, e.g. interpreting \ttt{IO k A} as the
\ttt{IO} monad in Haskell.  A trivial implementation of this interface
may run all computations in a single executable on a single machine.  A
more realistic implementation would require compiler support for (a)
compiling a single program to run on multiple hosts and (b) marshaling
and unmarshaling all values (the ML5 compiler implements both of these).
We also give a high-level abstract operational semantics for
computations in the companion code.


L5 satisfies the design goals of ML5: it ensures that resources are only
used by a computation running at the appropriate world, and it also
makes all communication explicit via \ttt{get}. 

The following example illustrates the use of the indexed types \ttt{Ref}
and \ttt{IO}, as well as the \ttt{get} operation on the monad:

\begin{code}
  update : (Ref server (IO server Unit))  →  IO client Unit  →  IO client Unit
  update l clicomp = get{server} (l := (get{client} clicomp)) 
\end{code}

\noindent This function takes a reference at the server that stores a
callback computation, which itself runs at the server, as well as a
computation that runs on the client, and produces a computation that
runs on the client.  This computation goes to the server (the outer
\ttt{get}), and updates the callback ref (\ttt{l}) to point to a
computation that goes back to the client and runs \ttt{clicomp} (the
inner \ttt{get}).  Omitting either \ttt{get} causes a type error,
preventing resources from being used at the wrong location.  In fact,
nothing about the code is specific to the worlds \ttt{client} and
\ttt{server}, so we can make it polymorphic in the worlds:

\begin{code}
  update' : (w1 : World) (w2 : World) → Ref w2 (IO w2 Unit) → IO w1 Unit → IO w1 Unit
  update' w1 w2 l comp1 = get{w2} (l := (get{w1} comp1))
\end{code}

\section{HL5}

A disadvantage of L5 is that types can be somewhat verbose, as they they
repeat the same world multiples times (e.g. both \ttt{client} and
\ttt{server} occur twice in the type of \ttt{update}).  A hybrid modal
type system, as in ML5, permits more concise specifications: one writes
a modal type \ttt{A}, which for the most part does not mention worlds,
and then interprets it relative to a world at the outside by writing
\ttt{A < w >}.  In this section, we define HL5, which is a hybrid type
system constructed on top of the above lax logic.  We define HL5 by
semantic embedding into L5 (which itself is just a library in Agda): we
define a syntax of HL5 types, and then an interpretation function
mapping HL5 types to functions from worlds to Agda \ttt{Set}s (or,
thinking constructively, predicates on worlds).  

\subsection{HL5 Types} The datatype of HL5 types is defined as follows
(note that Agda allows multiple datatype constructors per line):

\begin{code}
  data Type : Set where
    _⊃_ _∨_ : Type → Type → Type
    ∀₅ ∃₅ : (World → Type) → Type
    _at_ : Type → World → Type
    ◯    : Type → Type
    ref  : Type → Type
\end{code}
\ignore{
\begin{code}
    ↓₅ : (World → Type) → Type
    _∧_ : Type → Type → Type
    ⊤ ⊥ : Type
    list : Type → Type
\end{code}}
The types \cd{⊃} and \ttt{∨} represent functions and sums (the notation
\verb|_∨_| allows ∨ to be used infix).  The types \ttt{∀₅} and \cd{∃₅}
represent quantifiers over worlds.  Next, \cd{at} is a connective of
hybrid logic, which allows types to set the world at which the type is
interpreted.  Finally, \cd{◯} and \cd{ref} represent monadic
computations and references; note that they are \emph{not} indexed by a
world. Box and diamond can be defined using quantifiers and \cd{at}:
\cd{□ A = ∀₅ (λ w → A at w)} and \cd{◇ A = ∃₅ (λ w → A at w)}.

Below we define the interpretation function \ttt{A < w >}, which takes a
modal type and a world and produces an Agda \ttt{Set}.  Using modal
types, we can rewrite the type of \ttt{update} as follows:

\noagda \begin{code}
  update  : (((ref (◯ ⊤)) at server)  ⊃  ◯ ⊤  ⊃  ◯ ⊤) < client >
\end{code}
The type \ttt{((ref (◯ ⊤)) at server) ⊃ ◯ ⊤ ⊃ ◯ ⊤} says that
\ttt{update} takes a reference to a computation, located at the server,
along with a computation, and produces a computation. The Agda function
\ttt{A < w >} takes a \ttt{Type} and a \ttt{World} and produces a
\ttt{Set}; here, it is used to say that the whole type is interpreted
relative to the client.

The above polymorphic \ttt{update'} is typed as follows: 

\noagda \begin{code}
  update' : (□(∀₅ (λ w2 → ref (◯ ⊤) at w2  ⊃  ◯ ⊤ ⊃ ◯ ⊤))) <*>
\end{code}
where the postfix symbol \ttt{<*> : Type → Set} means that the
proposition is true in all worlds (i.e. \ttt{A <*>} means the Agda type
\ttt{\{w : World\} → A < w >}).  The outer \ttt{□} binds the "client"
world (i.e. the world for the computations in \ttt{◯ ⊤ ⊃ ◯ ⊤}); the
inner ∀₅ binds the "server" world (i.e. the world of the \ttt{ref (◯
⊤)}.

\ignore{
\begin{code}
  infix 100 _<_> -- more tightly than ::
  infixr 101 _⊃_ -- more tightly than [ ] 
  infixr 102 _∨_ -- more tightly than [ ] 
  infixr 103 _∧_ -- more tightly than [ ] 
  infix 104 _at_ -- more tightly than [ ] 
\end{code}}

\subsection{Interpretation}

We define an interpretation function \cd{A < w >} interpreting a type
\ttt{A} and a world \ttt{w} as an Agda classifier (a \ttt{Set}).  Then
the proof of an HL5 judgement \ttt{A1 < w1 > ... ⊢ C < w >} is an Agda
function of type \ttt{A1 < w1 > ... → C < w >}.  This interpretation is
a \emph{constructive Kripke semantics}---i.e. a Kripke semantics of an
intuitionistic modal logic in intuitionistic first-order logic, relative
to the Kripke structure given by the type
\ttt{World}.\footnote{Technically, we omit two of the pieces of a Kripke
structure: \ttt{World} is the set of states, but because we are
representing S5, we can elide the accessibility relation, and because we
do not have uninterpreted base \ttt{Type}s, we do not need an
interpretation of them.}  The interpretation is defined as as follows:

\begin{code}    
  _<_> : Type → World → Set
  (ref A) < w > = Ref w (A < w >)
  (◯ A) < w > = IO w (A < w >)
  (A at w') < _ > = A < w' >
  (A ⊃ B) < w > = A < w > → B < w >
  (A ∨ B) < w > = Either (A < w >) (B < w >)
  (∀₅ A) < w >   = (w' : World)  →  (A w') < w >
  (∃₅ A) < w >   = Σ \ (w' : World)  →  (A w') < w >
\end{code}
\ignore{
\begin{code}
  (A ∧ B) < w > = A < w > × B < w >
  ⊤ < w >       = Unit
  ⊥ < w >       = Void
  (list A) < w > = List (A < w >)
  (↓₅ A) < w >   = (A w) < w >
\end{code}}
The main action of the translations is to annotate \ttt{ref} and \ttt{◯}
with the world at which the type is being interpreted.  The hybrid
connective \cd{at} interprets its body at the specified world, ignoring
the current world.  Otherwise, the translation interprets each
connective as the corresponding Agda \ttt{Set}-former, with the
translation applied recursively.  Note
that the interpretation of \ttt{⊃} is simply \ttt{→}: when giving a
Kripke semantics for intuitionistic logic in a \emph{classical}
meta-language, it is necessary to interpret \ttt{A ⊃ B} as if it were
boxed, by quantifying over future worlds; but because our meta-language is
intuitionistic, this is not necessary here.  


The function \ttt{<*> : Type → Set} is defined to be the Agda set \cd{\{w
: World\} → A < w >}---\ttt{A} is true in all worlds, with the world
itself an implicit argument to the function.  Agda verifies that the
modal types of \ttt{update} and \ttt{update'} reduce to the explicit
types given above.

This formalization has several benefits: we can immediately use Agda to
program in the modal logic, and existing Agda code can be used at modal
types.  
For example, one can check that this definition validates all
of the rules of intuitionistic S5~\citep{simpson93thesis}, by
implementing the proof that the Simpson rules are sound for the Kripke
semantics.  The untethered elimination rule for disjunction and a
projective elimination rule for \ttt{at} are defined as follows:

\begin{code}
    casev : ∀ {A B C w w'} →  A ∨ B < w >
          → (A < w > → C < w' >)  →  (B < w > → C < w' >)
          → C < w' >
    casev (Inl e) b1 b2 = b1 e
    casev (Inr e) b1 b2 = b2 e

    unatv : ∀ {A w w'}  →  (A at w') < w >  →  A < w' >
    unatv x = x
\end{code}

Additionally, we can see that the return and bind operations on the
monad have their expected types at any world:

\begin{code}
    hl5ret : ∀ {A} →  A ⊃ ◯ A  <*>
    hl5ret = return

    hl5bind : ∀ {A B} →  ◯ A ⊃ (A ⊃ ◯ B) ⊃ ◯ B  <*>
    hl5bind =  _>>=_ 
\end{code}
The type of bind insists that all three ◯'s be at the same world:
sequencing tethers the premise to the conclusion.
 
Having considered return and bind, it is natural to ask what hybrid type
can we ascribe to the \ttt{get} operation on the indexed monad \ttt{IO w
A}. If we try defining

\noagda \begin{code}
    hl5get : ∀ {A w1 w2} → ((◯ A) at w1) ⊃ ((◯ A) at w2) <*>
\end{code}
we see that \ttt{hl5get} must transform \ttt{IO w1 (A < w1 >)} into \ttt{IO
w2 (A < w2 >)}.  L5 \ttt{get} satisfies this requirement if \ttt{A} is a
\emph{constant function} of its world argument, in which case \ttt{A < w1 >} is the same Agda 
\ttt{Set} as \ttt{A < w2 >}.

\ignore{
\begin{code}
  open Axioms

  data EqSet : (A B : Set) → Set1 where
    eq→ : ∀ {A1 A2 B1 B2 : Set} → EqSet B1 A1 → EqSet A2 B2 → EqSet (A1 → A2) (B1 → B2)
    eq× : ∀ {A1 A2 B1 B2 : Set} → EqSet A1 B1 → EqSet A2 B2 → EqSet (A1 × A2) (B1 × B2)
    eqEither : ∀ {A1 A2 B1 B2 : Set} → EqSet A1 B1 → EqSet A2 B2 → EqSet (Either A1 A2) (Either B1 B2)
    eqΠ : ∀ {A} {B1 B2 : A → Set} → ((x : A) → EqSet (B1 x) (B2 x)) → EqSet ((x : A) → B1 x) ((x : A) → B2 x)
    eqΣ : ∀ {A} {B1 B2 : A → Set} → ((x : A) → EqSet (B1 x) (B2 x)) → EqSet (Σ \ (x : A) → B1 x) (Σ \ (x : A) → B2 x)
    eqRefl : ∀ {A} → EqSet A A

  coerce : ∀ {A B} → EqSet A B → A → B
  coerce (eq→ y y') v = \ x → coerce y' (v (coerce y x))
  coerce (eqEither y y') (Inl y0) = Inl (coerce y y0)
  coerce (eqEither y y') (Inr y0) = Inr (coerce y' y0)
  coerce (eqΠ y) v = \ x → coerce (y x) (v x)
  coerce (eqΣ y) (x , y') = x , coerce (y x) y'
  coerce (eq× x y) (x1 , y1) = coerce x x1 , coerce y y1
  coerce eqRefl v = v
\end{code}}

We characterize constant modal types by the property that they yield the
same \ttt{Set} for any two arguments:

\begin{code}  
  Constant : Type → Set1
  Constant A = ∀ {w w'} → EqSet (A < w >) (A < w' >)
\end{code}
Here \ttt{EqSet} is an Agda relation expressing that the two \ttt{Set}s
are equal classifiers (in fact, we need a notion of equality that
compares the bodies of Π and Σ on all arguments, which we borrow from
OTT~\citep{altenkirch+07ott}); it is equipped with an operation
\ttt{coerce : ∀ \{A B\} → EqSet A B → A → B}.
It is simple to prove that \ttt{A at w} is constant, and that the
connectives ∨ ⊃ ∀₅ ∃₅ preserve constantness.  Neither \ttt{ref} nor ◯ is
constant, as their interpretation directly mentions the world.  Thus,
the constant types are those where all \ttt{ref}s and ◯'s are guarded by
an \ttt{at}.

\ignore{
\begin{code}
  iat : ∀ {A w} → Constant (A at w)
  iat = eqRefl
  
  i∨ : ∀ {A B} → Constant A → Constant B → Constant (A ∨ B)
  i∨ i1 i2 = eqEither i1 i2

  i∧ : ∀ {A B} → Constant A → Constant B → Constant (A ∧ B)
  i∧ i1 i2 = eq× i1 i2

  i⊃  : ∀ {A B} → Constant A → Constant B → Constant (A ⊃ B)
  i⊃ i1 i2 = eq→ i1 i2

  i∀  : ∀ {A} → ((w' : _) → Constant (A w')) → Constant (∀₅ A)
  i∀ iA = eqΠ (\ w → iA w)

  i∃  : ∀ {A} → ((w' : _) → Constant (A w')) → Constant (∃₅ A)
  i∃ iA = eqΣ (\ w → iA w)

  i⊤ : Constant ⊤
  i⊤ = eqRefl

  i⊥ : Constant ⊥
  i⊥ = eqRefl

  -- can't
  --   i◯  : ∀ {A} → Constant A → Constant (◯ A)
  --   i◯ ia = {!!}
\end{code}}

Now, we can ascribe \ttt{get} the following monadic type:
\begin{code}
  hl5get : ∀ {A w1 w2} → Constant A → ((◯ A) at w1) ⊃ ((◯ A) at w2) <*>
  hl5get con e = get e >>= \ v → return (coerce con v)
\end{code}
\ttt{hl5get} is equivalent to \ttt{get}, using the monad laws and the
fact that coercion based on a proof of \ttt{EqSet} is the identity (at
least up to extensional equality),

\section{ML5}

In this section, we give inductive definitions of the syntactic
apparatus of ML5: First, we define the syntax of types.  Next, we
represent programs using an \emph{intrinsic encoding}, which represents
only well-typed syntax (i.e. typing derivations or natural deduction
proofs).  Variables are represented as well-scoped de Bruijn
indices---pointers into a typing context, which is an explicit parameter
to the typing judgements.  The static semantics requires an auxiliary
definition of a \emph{mobility} judgement on types, which is defined
below.

\subsection{Types}

\begin{code}
  data Type5 : Set where
    _⇀_ _∨_  : Type5 → Type5 → Type5
    ∀₅ ∃₅ : (World → Type5) → Type5
    _at_ : Type5 → World → Type5
    ref  : Type5 → Type5
    ⌘   : (World → Type5) → Type5 
\end{code}
\ignore{
\begin{code}
    _∧_ : Type5 → Type5 → Type5
    ⊤ ⊥  : Type5
\end{code}}
The type \cd{⇀} represents partial functions, and the type \cd{∨}
represents sums.  The types \ttt{∀₅} \cd{∃₅} cd{at} and \cd{ref} are the
same as in HL5.  There is an additional type constructor "shamrock",
rendered here as ⌘, which is a □-like modality, but its rules in the ML5
proof theory are subtly different than the rules for \ttt{∀₅ (λ w → A at
w)}.

We represent types with free world variables as Agda functions from
worlds to types.  This is permissible because \ttt{World} is defined
prior to type (i.e. we are using Weak HOAS\cite{despeyroux+95hoascoq,bucalo+06contexts}).
If \ttt{World} is chosen to be a base type in Agda, then it adequately
represents ML5 types as in \citet{murphy08thesis}.  If instead
\ttt{World} is chosen to be an inductive type, this representation
yields a language with type-level case analysis over worlds---i.e. one
could define types whose structure varies depending on their world, such
as \ttt{∃₅ (λ w → if w = server then nat else bool)}; we leave an
exploration of the practical uses of this alternative to future work.

\ignore{
\begin{code}
  infix 100 _[_] -- more tightly than ::
  infix 10 _⊢_ -- pretty loose
  infix 101 _⇀_ -- more tightly than [ ] 
  infix 102 _∨_ -- more tightly than [ ] 
  infix 103 _∧_ -- more tightly than [ ] 
  infix 104 _at_ -- more tightly than [ ] 
\end{code}}

\subsection{Mobility}

ML5's notion of mobility identifies constant functions, analogously to
the \ttt{Constant} relation on HL5 types above.

\begin{code}
  data Mobile5 : Type5 → Set where
    mat5 : ∀ {A w} → Mobile5 (A at w)
    m⌘5 : ∀ {A} → Mobile5 (⌘ A)
    ---- no rule for ⇀ or refs 
    m∨5  : ∀ {A B} → Mobile5 A → Mobile5 B → Mobile5 (A ∨ B)
    m∀5  : ∀ {A} → ((w' : _) → Mobile5 (A w')) → Mobile5 (∀₅ A)
    m∃5  : ∀ {A} → ((w' : _) → Mobile5 (A w')) → Mobile5 (∃₅ A)
\end{code}
\ignore{
\begin{code}
    m∧5  : ∀ {A B} → Mobile5 A → Mobile5 B → Mobile5 (A ∧ B)
    m⊤5  : Mobile5 ⊤
    m⊥5  : Mobile5 ⊥
\end{code}}
To foreshadow the interpretation: ⌘ is always mobile, essentially
because ∀₅ (λ w → (A w) at w) is constant.  Functions are never mobile,
because they hide a ◯ in their domain, and ◯ is not constant.

\subsection{Typing judgements}

ML5 judgements have the form 
\cd{Γ ⊢ e : A [ w ]} and 
\cd{Γ ⊢ v :: A [ w ]}.
These judgements mean that the expression \ttt{e} and the value \ttt{v}
are well-formed with modal type \ttt{A} at world \ttt{w}.
Here Γ contains assumptions \ttt{x:A[w]} and \ttt{u $\sim$ w.A}.  
The former, a \emph{value hypothesis}, means that \ttt{x} stands for a
value of type \ttt{A} at world \ttt{w}.  The latter, a \emph{valid
hypothesis}, means that \ttt{u} stands for a value of type \ttt{A} that
makes sense at all worlds (\ttt{w} is bound in \ttt{A}).  
In the operational semantics, we will substitute a proof of the judgement
\ttt{w:world ⊢ v :: (A w)[ w ]} for a valid variable.  

We combine both of the above judgements into one Agda type, defining a
relation \ttt{Γ ⊢ γ}.  Here Γ is a list of hypotheses, which are either
\ttt{value (A [ w ])} or \ttt{valid (w.A)} and γ is a conclusion, which
is either \ttt{exp (A [ w ])} (for expressions) or \ttt{value (A [ w ])}
(for values).  As in the syntax of types, \ttt{w.A} is represented by an
Agda function from worlds to types.  \ttt{⊢} binds more tightly than →,
so we can write e.g. \ttt{Γ ⊢ C1 → Γ ⊢ C2}  for \ttt{(Γ ⊢ C1) → (Γ ⊢
C2)}.

We define the notation \ttt{A [ w ]} to mean the pair of \ttt{A} and
\ttt{w}, and we define sum types for hypotheses and conclusions as
follows:

\ignore{
\begin{code}
  LocType : Set 
  LocType = Type5 × World

  _[_] : Type5 → World → LocType
  _[_] = _,_ 
\end{code}}

\begin{code}
  data Hyp : Set where
    value : (Type5 × World) → Hyp
    valid : (World → Type5) → Hyp

  data Conc : Set where
    exp  : (Type5 × World) → Conc
    value : (Type5 × World) → Conc

  Ctx = List Hyp
\end{code}

Term variables \ttt{x} and \ttt{u} are represented by
well-scoped de Bruijn indices---pointers into Γ:

\noagda \begin{code}
data _∈_ {A : Set} : A → List A → Set where
 i0 : {α : A}   {Γ : List A} → α ∈ (α :: Γ)
 iS : {α α' : A} {Γ : List A} → α ∈ Γ → α ∈ (α' :: Γ)
\end{code}

The typing rules are defined in Figure~\ref{fig:ml5-typing}.  The first
rule says that \ttt{(▹ x)} is a value if \ttt{x} is a de Bruijn index
for a value assumption in Γ.  The next rule represents the application
of a valid assumption in Γ to a world \ttt{w}; we use propositional
equality \ttt{Id} to state that the conclusion type is \ttt{A w} because
otherwise the higher-order conclusion blocks pattern-matching.
\ttt{lam} takes an expression in an extended context to a value of
function type; this rule expresses the idea that a function of type
\ttt{A ⇀ B} in a world \ttt{w} is an expression of type \ttt{B} at
\ttt{w}, hypothetically in a variable standing for a value of type
\ttt{A} at \ttt{w}.  Sums are introduced by commuting the world with the
connective.  \ttt{hold} switches worlds to introduce an \ttt{at};
\ttt{wlam} and \ttt{wpair} introduce quantifiers in the usual way; the
world in the conclusion does not change (also, note the parentheses,
which in Agda's notation for datatype constructors are the only
difference between the two rules: the premise of \ttt{wlam} is a
function, whereas \ttt{wpair} has two premises).  However, note that the
body of a \ttt{wlam} is a value, not an expression---ML5 has a value
restriction on world quantification, to support type inference in the
style of ML.  \ttt{sham} both quantifies over a new world and switches
the world of the conclusion---indeed, this is the derived intro rule
that one would expect for \ttt{⌘ A = ∀₅ (λ w → (A w) at w)}.

\ttt{val} injects values into expressions, whereas \ttt{let} sequences
the evaluation of two expressions.  \ttt{put} runs an expression of
mobile type and then binds the resulting value as a valid assumption.
\ttt{app} and \ttt{wapp} eliminate functions and universals, with the
world along for the ride.  The remaining rules are
pattern-matching-style elimination rules. In these rules, the world in
the conclusion \ttt{C [ w ]} is \emph{tethered} to the world in the
principal premise, as discussed in the introduction.

This system is the ML5 internal language described in Section 4.3 of
\citep{murphy08thesis} with one minor change: we have eliminated the
syntactic class of valid values, which allowed validity to appear as a
conclusion as well as as a hypothesis---valid values are unnecessary
because validity is invertible on the right.  An alternate formalization
with valid values is included in the companion code.

\begin{figure}
\begin{footnotesize}
\begin{multicols}{2}
\ignore{
\begin{code}
  _,,_ : ∀ {A} → List A → A → List A
  Γ ,, A = A :: Γ
  infixl 10 _,,_
\end{code}}
\begin{code}
  data _⊢_ (Γ : Ctx) : Conc → Set where

    ---- values
    ▹ : ∀ {A w} 
     → value (A [ w ]) ∈ Γ 
     → Γ ⊢ value (A [ w ])

    ▹v : ∀ {w A C} 
     → valid A ∈ Γ  →  Id C (A w) 
     → Γ ⊢ value (C [ w ])

    lam : ∀ {A B w} 
     → (Γ ,, value (A [ w ])) ⊢ exp (B [ w ])
     → Γ ⊢ value (A ⇀ B [ w ])

\end{code}
\ignore{
\begin{code}
    pair : ∀ {A B w} 
     → Γ ⊢ value (A [ w ])
     → Γ ⊢ value (B [ w ]) 
     → Γ ⊢ value (A ∧ B [ w ])

    mt : ∀ {w} 
     → Γ ⊢ value (⊤ [ w ])
\end{code}}
\begin{code}
    inl : ∀ {A B w} 
     → Γ ⊢ value (A [ w ])
     → Γ ⊢ value (A ∨ B [ w ])

    inr : ∀ {A B w} 
     → Γ ⊢ value (B [ w ])
     → Γ ⊢ value (A ∨ B [ w ])

    hold : ∀ {A w w'}
     → Γ ⊢ value (A [ w' ])
     → Γ ⊢ value ((A at w') [ w ])

    wlam : ∀ {A w}
     → ((w' : _) → Γ ⊢ value ((A w') [ w ]))
     → Γ ⊢ value (∀₅ A [ w ]) 

    wpair : ∀ {A w} 
     → (w' : World) 
     → Γ ⊢ value ((A w') [ w ]) 
     → Γ ⊢ value ((∃₅ A) [ w ])

    sham : ∀ {A w'} 
     → ((w : _) → Γ ⊢ value (A w [ w ])) 
     → Γ ⊢ value (⌘ A [ w' ])

    ----- expressions

    val : ∀ {L} 
     → Γ ⊢ value L 
     → Γ ⊢ exp L

    lete : ∀ {A C w} 
     → Γ ⊢ exp (A [ w ]) 
     → (Γ ,, value (A [ w ])) ⊢ exp (C [ w ])
     → Γ ⊢ exp (C [ w ])

    get5 : ∀ {A w w'} 
      → Γ ⊢ exp (A [ w' ])  →  Mobile5 A 
      → Γ ⊢ exp (A [ w ])

    put : ∀ {A C w} 
      → Γ ⊢ exp (A [ w ])  →  Mobile5 A 
      → (Γ ,, valid (\ _ → A)) ⊢ exp (C [ w ])
      → Γ ⊢ exp (C [ w ])

    app : ∀ {A B w} 
      → Γ ⊢ exp (A ⇀ B [ w ])  
      → Γ ⊢ exp (A [ w ])
      → Γ ⊢ exp (B [ w ])

    wapp : ∀ {A w} 
      → Γ ⊢ exp (∀₅ A [ w ])  
      → (w' : World) 
      → Γ ⊢ exp ((A w') [ w ])

    case : ∀ {A B C w} 
      → Γ ⊢ exp (A ∨ B [ w ])  
      → (Γ ,, value (A [ w ]) ⊢ exp (C [ w ])) 
      → (Γ ,, value (B [ w ])  ⊢ exp (C [ w ])) 
      → Γ ⊢ exp (C [ w ])
\end{code}
\ignore{
\begin{code}
    split : ∀ {A B C w} 
      → Γ ⊢ exp (A ∧ B [ w ]) 
      → Γ ,, value (A [ w ]) ,, value (B [ w ]) 
          ⊢ exp (C [ w ])
      → Γ ⊢ exp (C [ w ])

    abort : ∀ {w C} 
      → Γ ⊢ exp (⊥ [ w ])
      → Γ ⊢ exp (C [ w ])
\end{code}}
\begin{code}
    wunpack : ∀ {A w C} 
      → Γ ⊢ exp ((∃₅ A) [ w ])
      → ((w' : _) → 
         Γ ,, value ((A w') [ w ]) ⊢ exp (C [ w ]))
      → Γ ⊢ exp (C [ w ])

    leta : ∀ {A w w' C} 
      → Γ ⊢ exp ((A at w') [ w ])  
      → (Γ ,, value (A [ w' ]) ⊢ exp (C [ w ])) 
      → Γ ⊢ exp (C [ w ])

    lets : ∀ {A w C} 
      → Γ ⊢ exp (⌘ A [ w ])
      → (Γ ,, valid A ⊢ exp (C [ w ])) 
      → Γ ⊢ exp (C [ w ])
\end{code}
\end{multicols}
\end{footnotesize}

\caption{ML5 Static Semantics}
\label{fig:ml5-typing}

\end{figure}

\ignore{
\begin{code}
open import lib.Prelude hiding ([_])
open import agda5.Agda5 

open import agda5.ML5Valid

module agda5.ML5Valid-Semantics where
\end{code}}

\section{Semantics}

Our semantics explains the meaning of an ML5 program by translation into
HL5.  As discussed above, our translation clarifies the essential
difference between the role that the world plays in the judgements
\ttt{value (A [ w ])} and \ttt{exp (A [ w ])}: a value hypothesis or
conclusion \ttt{value (A [ w ])} is interpreted as a \emph{pure term} of
HL5 type \ttt{(eff A) < w >}, where \ttt{eff A} is a monadic translation
of \ttt{A}.  Thus, the role of the world \ttt{w} is only to describe
where the resources in \ttt{A} may be used and where the computations
must be run.  On the other hand, an expression \ttt{exp (A [ w ])} is
interpreted as a \emph{monadic computation} of type \ttt{(◯ (eff A)) < w
>}, so the world determines both the site at which the effectful
computation must be run and where the resources/computations in \ttt{A}
may be used.

\subsection{Type Translation}

The type translation from ML5 types to HL5 translates ⌘ to HL5 □, adds a ◯
to the codomain of ⇀ (as in any monadic
translation~\citep{moggi91monad}), and is defined compositionally otherwise.  The
ML5 connective \ttt{∀₅} has a value-restriction, so there is no ◯
inserted in its body.  Note that Agda allows datatype constructors to be
overloaded, so we can have both ML5 types and HL5 types in scope at once.

\begin{code}
  eff : Type5 → Type
  eff (A ⇀ B) = eff A  ⊃  ◯(eff B)
  eff (A ∨ B) = eff A ∨ eff B
  eff (∀₅ A) = ∀₅ (λ w → (eff (A w)))
  eff (∃₅ A) = ∃₅ (λ w → eff (A w))
  eff (A at w) = (eff A) at w 
  eff (ref A) = ref (eff A)
  eff (⌘ A) = ∀₅ (λ w → ((eff (A w)) at w))
\end{code}  
\ignore{
\begin{code}
  eff (A ∧ B) = eff A ∧ eff B
  eff ⊤ = ⊤
  eff ⊥ = ⊥
\end{code}}

A simple induction verifies that mobile ML5 types translate to constant
HL5 types:

\begin{code}
  eff-mobile : ∀ {A} → Mobile5 A → Constant (eff A)
\end{code}

\ignore{
\begin{code}
  eff-mobile (mat5{A}{w}) {w1}{w2} = iat{eff A}{w}{w1}{w2}
  eff-mobile (m∧5{A}{B} y y') = i∧ {eff A}{eff B} (eff-mobile y) (eff-mobile y')
  eff-mobile (m∨5{A}{B} y y') = i∨ {eff A}{eff B} (eff-mobile y) (eff-mobile y')
  eff-mobile m⊤5 {w1}{w2} = i⊤{w1}{w2}
  eff-mobile m⊥5 {w1}{w2} = i⊥{w1}{w2}
  eff-mobile (m∀5{A} y) = i∀{\ w -> eff (A w)} \ w → (eff-mobile (y w))
  eff-mobile (m∃5{A} y) = i∃{\ w -> eff (A w)} \ w → (eff-mobile (y w))
  eff-mobile (m⌘5{A}) {w1}{w2} = (i∀{\ w -> eff (A w) at w} \w → iat{eff (A w)}{w}{w1}{w2}) {w1}{w2}
\end{code}}

\subsection{Term translation}

Next, we interpret hypotheses and conclusions.  Values and expressions
are interpreted as described above (note that for \ttt{L = A [ w ]},
\ttt{fst L = A} and \ttt{snd L = w}).  A valid hypothesis is interpreted
as the Agda set \ttt{(w : World) → (interp-hyp (value (A w [ w ])))},
though to appease the termination checker we have to unroll the
definiton of \ttt{interp-hyp}.

\begin{code}
  interp-hyp : Hyp → Set
  interp-hyp (value L) = (eff (fst L)) < (snd L) >
  interp-hyp (valid A) = (w : World) → (eff (A w)) < w >

  interp-conc : Conc → Set
  interp-conc (value L) = interp-hyp (value L) 
  interp-conc (exp L) = (◯ (eff (fst L))) < (snd L) >
\end{code}

Next, we interpret the syntax (Figure~\ref{fig:interp}).  The Agda type
\ttt{Everywhere P xs} represents a list with one element of type \ttt{P
x} for each element \ttt{x} in \ttt{xs}; we use this to represent a
substitution of semantic values for syntactic variables.  The usual
propositional connectives are interpreted in the standard way, by
sequencing evaluation if necessary and then applying the corresponding
Agda introduction or elimination form.  Subexpressions that are under
bound variables are interpreted in an extended context (e.g. \ttt{e} in
\ttt{lam e}).  \ttt{ListM.EW.there} looks up a de Bruijn index into Γ in
a substitution of type \ttt{Everywhere P Γ}.  

\ttt{sham} is interpreted as an Agda function, which is applied in the
translation of the \ttt{▹v} rule for using valid variables.  There are
no Agda term constructors for \ttt{at}, so the translation simply
proceeds inductively.  The quantifiers are interpreted by supplying the
world arguments in the syntax, rather than by extending the substitution
(recall that the proof terms \ttt{wlam} and \ttt{wunpack} contain Agda
functions).  Recall that the body of \ttt{wlam} is already a value, so
no further evaluation is necessary.

\ttt{val} and \ttt{lete} are interpreted directly as return and bind.
\ttt{get5} is translated as a \ttt{get} in the target, followed by a
coercion by the mobility proof.  \ttt{put e} extends the substitution
with the value of \ttt{e}, applying the mobility proof to the value so
it can be used at any other world.

Agda verifies that the interpretation is type correct and total,
establishing that the interpretation is total and type-preserving.  

\begin{figure}[t]
\begin{code}
  eval : ∀ {Γ L} → Γ ⊢ L → Everywhere interp-hyp Γ → interp-conc L
  ---- usual connectives
  eval (▹ x) σ = (ListM.EW.there σ x)
  eval (lam e) σ = λ x → eval e (x E:: σ)
  eval (app e1 e2) σ =  (eval e1 σ) >>= λ f → (eval e2 σ) >>= λ a → f a 
  eval (inl v) σ = Inl (eval v σ)
  eval (inr v) σ = Inr (eval v σ)
  eval (case e e1 e2) σ = eval e σ >>= docase where
    docase : Either _ _ → _
    docase (Inl x) = eval e1 (x E:: σ)
    docase (Inr y) = eval e2 (y E:: σ)
  ---- ⌘ 
  eval (▹v{w} x Refl) σ =  (ListM.EW.there σ x) w
  eval (sham v) σ =  λ w' → (eval (v w') σ)
  eval (lets e e') σ = eval e σ >>= λ x → eval e' (x E:: σ)
  ---- at
  eval (hold v) σ = eval v σ
  eval (leta e e') σ = eval e σ >>= λ x → eval e' (x E:: σ)
  ---- ∀ and ∃
  eval (wlam v) σ = λ w → eval (v w) σ
  eval (wapp e w) σ = eval e σ >>= λ f → return (f w)
  eval (wpair w v) σ = w , eval v σ
  eval (wunpack e e') σ = eval e σ >>= λ p → eval (e' (fst p)) (snd p E:: σ)
  --- monad operations and world movement
  eval (val v) σ = return (eval v σ)
  eval (lete e1 e2) σ = eval e1 σ >>= λ x → eval e2 (x E:: σ)
  eval (get5{A} e mob) σ =  get (eval e σ) >>= λ v → return (coerce (eff-mobile mob) v) 
  eval (put e mob e') σ = eval e σ >>= λ v → 
                           eval e' ((λ w' -> coerce (eff-mobile mob) v) E:: σ)
\end{code}
\ignore{
\begin{code}
  eval (pair v1 v2) σ = eval v1 σ , eval v2 σ
  eval (split e e1) σ = (eval e σ) >>= λ p → eval e1 (snd p E:: fst p E:: σ) 
  eval mt σ = <>
  eval (abort e) σ = (eval e σ) >>= Sums.abort

  evalc : ∀ {L} → [] ⊢ L → interp-conc L
  evalc e = eval e E[]
\end{code}}

\caption{Interpretation of the syntax}
\label{fig:interp}
\smallskip
\hrule
\end{figure}

\section{Revised ML5}

In this section, we simplify and generalize the ML5 source language
based on the above semantics.  First, our analysis has shown that
\ttt{Mobile5 A} really means that \ttt{A [ w ]} and \ttt{A [ w' ]} are
equal types for any \ttt{w} and \ttt{w'}.  We can internalize this
principle as a value coercion \ttt{vshift} in
Figure~\ref{fig:ml5-revised}.  This makes it possible to implement some
additional programs without communication.  For example, consider a
function

\noagda \begin{code}
move : ∀ {A w w'} →  Mobile5 A  →  (value (A [ w ]) :: []) ⊢ exp (A [ w' ])
move m =  (get5 (val (▹ i0)) m)
\end{code}
Operationally, this is quite inefficient: it sends the value of the
variable from \ttt{w'} to \ttt{w}, as the environment of the \ttt{val (▹
i0)} closure, and \ttt{w} then returns this value back to \ttt{w'}.
This whole process is unnecessary, as the value was available at
\ttt{w'} to begin with!  However, it does not seem possible to implement
this function without communication in ML5.
In our revised ML5, as in HL5, it can be implemented as a simple type
coercion:

\noagda \begin{code}
move : ∀ {A w w'} →  Mobile5 A  →  (value (A [ w ]) :: []) ⊢ exp (A [ w' ])
move m = val (vshift (▹ i0) m)
\end{code}

Second, in HL5, the worlding of values always "gets out of the way"
because it commutes with type constructors down to \ttt{ref} and
\ttt{◯}.  In Figure~\ref{fig:ml5-revised}, we add these non-local
elimination rules for values of each connective to the source.  For
example, we add the untethered case rule for values of sum type and a
projective elimination rule for \ttt{at} and ∀₅ values.  

We could additionally add elimination rules like \ttt{case},
\ttt{split}, etc. to the syntactic class of "values"---i.e. we could
admit that we are really dealing with a syntactic class of pure terms:
\noagda \begin{code}
 casev/val : ∀ {A B C w' w} → Γ ⊢ value (A ∨ B [ w ])  
   → (Γ ,, value (A [ w ]) ⊢ val (C [ w' ])) → (Γ ,, value (B [ w ]) ⊢ val (C [ w' ])) 
   → Γ ⊢ val (C [ w' ])
\end{code}
However, in an operational semantics where worlds and types are erased
at run-time, \ttt{wappv} and \ttt{unatv} will not create any real
redexes at run-time, whereas \ttt{casev/val} will.  Thus a reasonable
design choice for ML5 would be to allow \ttt{wappv} and \ttt{unatv} but
not the corresponding \ttt{case} rule.

\begin{figure}[t]
\noindent New value rules:
\begin{code}
    vshift : ∀ {A w w'} → Γ ⊢ value (A [ w ]) →  Mobile5 A → Γ ⊢ value (A [ w' ])
    wappv : ∀ {A w} → Γ ⊢ value (∀₅ A [ w ])  →  (w' : World) → Γ ⊢ value ((A w') [ w ])
    unatv : ∀ {A w w'} → Γ ⊢ value ((A at w') [ w ]) → Γ ⊢ value (A [ w' ])
\end{code}
\ignore{
\begin{code}
    ----- expressions

    val : ∀ {L} 
     → Γ ⊢ value L 
     → Γ ⊢ exp L

    lete : ∀ {A C w} 
     → Γ ⊢ exp (A [ w ]) 
     → (Γ ,, value (A [ w ])) ⊢ exp (C [ w ])
     → Γ ⊢ exp (C [ w ])

    -- or just make it admissible?
    letv : ∀ {L J} 
     → Γ ⊢ value L
     → (value L :: Γ) ⊢ J
     → Γ ⊢ J 

    get5 : ∀ {A w w'} 
      → Γ ⊢ exp (A [ w' ])  →  Mobile5 A 
      → Γ ⊢ exp (A [ w ])

    app : ∀ {A B w} 
      → Γ ⊢ exp (A ⊃ B [ w ])  →  Γ ⊢ exp (A [ w ])
      → Γ ⊢ exp (B [ w ])

    splitv : ∀ {A B L w} → Γ ⊢ value (A ∧ B [ w ]) 
      → Γ ,, value (A [ w ]) ,, value (B [ w ]) ⊢ exp L → Γ ⊢ exp L

    abortv : ∀ {w L} → Γ ⊢ value (⊥ [ w ]) → Γ ⊢ exp L
\end{code}}
\noindent New expression rules:
\begin{code}
    casev : ∀ {A B L w} → Γ ⊢ value (A ∨ B [ w ])  
      → (Γ ,, value (A [ w ]) ⊢ exp L) → (Γ ,, value (B [ w ])  ⊢ exp L) 
      → Γ ⊢ exp L
    wunpackv : ∀ {A w L} → Γ ⊢ value ((∃₅ A) [ w ]) 
             → ((w' : _) → Γ ,, value ((A w') [ w ]) ⊢ exp L) → Γ ⊢ exp L
\end{code}

Remove rules \ttt{put}, \ttt{▹v}, \ttt{lets}, \ttt{sham}, \ttt{wapp},
\ttt{leta}, \ttt{case}, \ttt{wunpack}.
\caption{Revised ML5}
\label{fig:ml5-revised}
\smallskip
\hrule
\end{figure}

It is simple to extend the semantics to these new constructs, as they
were derived from it:

\begin{code}
    eval (vshift e m) σ = coerce (eff-mobile m) (eval e σ)
    eval (wappv e w') σ = eval e σ w'
    eval (unatv e ) σ = eval e σ
    eval (casev v e1 e2) σ = docase (eval v σ) where
      docase : Either _ _ → _
      docase (Inl x) = eval e1 (x E:: σ)
      docase (Inr y) = eval e2 (y E:: σ)
    eval (wunpackv v e') σ = let p = eval v σ in eval (e' (fst p)) (snd p E:: σ)
\end{code}

\ignore{\begin{code}
    evalc : ∀ {L} → [] ⊢ L → interp-conc L
    evalc e = eval e E[] 
\end{code}}

\paragraph{Derived Forms}

Once we have added the aforementioned rules, we can remove rules
\ttt{put}, \ttt{▹v}, \ttt{lets}, \ttt{sham}, \ttt{wapp}, \ttt{leta},
\ttt{case}, \ttt{wunpack}, which become derivable.  Shamrock is defined
using \ttt{∀₅} and \ttt{at} in the straightforward way: \ttt{⌘ A = ∀₅ (λ
w → A w at w)}. Validity is defined as a value assumption of shamrock
type: \ttt{valid A = value ((⌘ A) [ dummy ])}, where \ttt{dummy} is any
arbitrary world---\ttt{⌘ A} is mobile, and therefore constant,
but because all value assumptions are worlded, we must pick one.

\ignore{
\begin{code}
  module Derived where
   open ListM.Subset 

   postulate
     weaken : ∀ {Γ Γ' J} → Γ ⊆ Γ' → Γ ⊢ J → Γ' ⊢ J

   dummy : World
   dummy = client

   ⌘ : (World → Type5) → Type5
   ⌘ A = ∀₅ (\ w → A w at w)
 
   valid : (World → Type5) → Hyp
   valid A = value ((⌘ A) [ dummy ])
\end{code}}

The term \ttt{▹v}, which represents a use of a valid hypothesis, is
derived by eliminating a shamrock.  The term \ttt{put} is derived by
introducing a shamrock, using \ttt{vshift} to retype the assumption in
the body---we use \ttt{letv} to refer to the substitution principle
plugging a value in for a variable, and we use \ttt{weaken} for
weakening in the de Bruijn syntax.

\begin{code} 
   ▹v : ∀ {Γ A C w} → valid A ∈ Γ  →  Id C (A w) → Γ ⊢ value (C [ w ])
   ▹v i Refl = unatv (wappv (▹ i) _)
   
   put : ∀ {Γ A C w} → Γ ⊢ exp (A [ w ])  →  Mobile5 A 
       → (Γ ,, valid (\ _ → A)) ⊢ exp (C [ w ]) → Γ ⊢ exp (C [ w ])
   put e m e' = lete e (letv (wlam (\ w' → hold (vshift (▹ i0) m))) 
                             (weaken (extend⊆ iS) e'))
 \end{code}
The remaining derivabilities show that the old rules for the
connectives are derivable using the new ones.

\section{Related Work}

\citet{murphy08thesis} describes ML5 and related languages, such as work
by \citet{jiawalker04modal}.  

\citet{crary98thesis,benke+03univeres,altenkirch03generic,chlipala07certifying}
describe other uses of universes and semantic embeddings in type theory,
though they do not consider embedding a modal type system.  We have used
the same technique for embedding a hybrid type system in Agda in
previous work~\citep{lh09unibind}.  Our technique is quite similar to
that of \citet{allen98notation}, who defines modal types as display
forms for NuPRL types.  The technical difference is that Allen considers
the modal types simply as notation, whereas in our approach the modal
types are data, equipped with a translation to meta-language types.
This shows how to achieve similar convenience of notation, without
requiring separate display-form facilities.  \citet{avron+98modal}
consider representations of modal logics in LF~\citep{hhp93lf}, some of
which use world-indexed judgements to track scoping.  We also use a
world-indexed type family \ttt{A < w >}, but this relation is defined
semantically (by interpretation into Agda) rather than syntactically (by
inference rules).

At the core, our interpretation reduces ML5 to L5, a language with an
indexed monad \ttt{IO w A} of computations at a place.  Indexed monads
have been studied in a variety of previous work, including
\citet{abadi+99dcc,atkey09parametrized,nanevski+06separation,russo+08informationflow}.
However, to interpret ML5, we require the programming language to
provide quantification over the indices to the monad, which
DCC~\citep{abadi+99dcc}, for example, does not provide.  It would be
interesting to adapt our work to these other settings, using a modal
logic to manage the indices to the monad.

\section{Conclusion}
\label{sec:soundness}

While we have used Agda for our development, we conjecture that the work
described in this paper could be carried out with similar effort in
Coq~\citep{inria06coqmanual}, as we have not used very complicated
dependent pattern matching.  However, in future work we would like to
embed proof-based access control following PCML5~\citep{avijit10pcml5},
which will require a modal universe with dependent types. Dependently
typed universes are easiest to represent using
induction-recursion~\citep{dybjersetzer01indrec}, which Agda supports
but Coq does not.

In future work, we also plan to complete a proof that the operational
semantics of ML5 are sound for the denotational semantics. We have
formalized the operational semantics of $\lambda_5$ and an operational
semantics for computations.
%
%
%
%
%
We have also proved soundness, assuming a standard compositionality lemma
(substitution of interpretations is the interpretation of the
substitution), which we are in the process of formalizing.  β-reduction
in Agda validates the β-steps for functions, sums, etc. in the source.
Because compositionality is really a property only of the binding
structure of the language and the semantics, not of the particular
language constructs, it should be possible to
implement compositionality in a datatype-generic manner, as in
\citet{chlipala07certifying}.  We also leave the question of full
abstraction to future work.

\paragraph{Acknowledgements}  We thank Jason Reed and Rob Simmons for
discussions about this article, and the anonymous reviewers for their
helpful comments.    

\setlength{\bibsep}{1pt} 
{ \small
\bibliographystyle{abbrvnat}
\bibliography{drl-common/cs}
}

\end{document}